\documentclass[%
 reprint,
nofootinbib,
nobibnotes,
 amsmath,amssymb,
aps,prl,twocolumn,showpacs,superscriptaddress
]{revtex4-2}

\usepackage{graphicx}
\usepackage{dcolumn}
\usepackage{bm}

\usepackage{booktabs}
\usepackage{multirow}
\usepackage{amsmath}

\begin{document}

\preprint{APS/123-QED}

\setcounter{secnumdepth}{2}
\renewcommand{\thesection}{\Roman{section}}
\renewcommand{\thesubsection}{\Alph{subsection}}

\title{Isoscalar Giant Resonances in Highly-Deformed $^{172}$Yb}

\author{K. Khokhar} 
\affiliation{Department of Physics, IIT-ISM Dhanbad, Jharkhand 826004, India}
\author{S. Bagchi}
\thanks{Contact author: \texttt{sbagchi@iitism.ac.in}}
\affiliation{Department of Physics, IIT-ISM Dhanbad, Jharkhand 826004, India}
\author{Y. Niu}
\affiliation{Frontiers Science Center for Rare isotopes, Lanzhou University, Lanzhou 730000, China}
\affiliation{School of Nuclear Science and Technology, Lanzhou University, Lanzhou 730000, China}
\author{C. Chen}
\affiliation{Frontiers Science Center for Rare isotopes, Lanzhou University, Lanzhou 730000, China}
\affiliation{School of Nuclear Science and Technology, Lanzhou University, Lanzhou 730000, China}
\author{M. N. Harakeh}
\affiliation{Nuclear Energy Group, ESRIG, University of Groningen, 9747 AA Groningen, The Netherlands}
\author{M. Abdullah}
\affiliation{Department of Physics, IIT-ISM Dhanbad, Jharkhand 826004, India}
\author{H. Akimune}
\affiliation{Department of Physics, Konan University, Hyogo 658-8501, Japan}  
\author{D. Das}
\thanks{Present address: TU Darmstadt and GSI Helmholtzzentrum, Germany.}
\affiliation{Department of Physics, IIT-ISM Dhanbad, Jharkhand 826004, India}
\author{T. Doi}
\affiliation{Department of Physics, Kyoto University, Kyoto 606-8502, Japan}
\author{L. M. Donaldson}
\affiliation{iThemba LABS, Somerset West, 7129, South Africa}
\author{Y. Fujikawa}
\affiliation{Department of Physics, Kyoto University, Kyoto 606-8502, Japan}
\author{M. Fujiwara}
\affiliation{Research Center for Nuclear Physics, Osaka University, Osaka 567-0047, Japan}
\author{T. Furuno}
\affiliation{Department of Physics, Osaka University, Toyonaka, Osaka 560-0043, Japan}
\affiliation{Department of Applied Physics, University of Fukui, Fukui 910-8507, Japan,}
\author{U. Garg}
\affiliation{Department of Physics, University of Notre Dame, Notre Dame, Indiana 46556, USA}
\author{Y. K. Gupta}
\affiliation{Nuclear Physics Division, Bhabha Atomic Research Centre, Mumbai 400085, India}
\affiliation{Homi Bhabha National Institute, Anushaktinagar, Mumbai 400094, India}
\author{K. B. Howard}
\affiliation{Department of Physics, University of Notre Dame, Notre Dame, Indiana 46556, USA}
\author{Y. Hijikata}
\affiliation{Department of Physics, Kyoto University, Kyoto 606-8502, Japan}
\affiliation{RIKEN Nishina Center for Accelerator-Based Science, Wako, Saitama 351-0198, Japan}
\author{K. Inaba}
\affiliation{Department of Physics, Kyoto University, Kyoto 606-8502, Japan}
\author{S. Ishida}
\affiliation{Cyclotron and Radioisotope Center, Tohoku University, Sendai 980-8578, Japan}
\author{M. Itoh}
\affiliation{Cyclotron and Radioisotope Center, Tohoku University, Sendai 980-8578, Japan}
\author{N. Kalantar-Nayestanaki}
\affiliation{Nuclear Energy Group, ESRIG, University of Groningen, 9747 AA Groningen, The Netherlands}
\author{D. Kar}
\affiliation{Department of Physics, IIT-ISM Dhanbad, Jharkhand 826004, India}
\author{T. Kawabata}
\affiliation{Department of Physics, Osaka University, Toyonaka, Osaka 560-0043, Japan}
\author{S. Kawashima}
\affiliation{Department of Physics, Konan University, Hyogo 658-8501, Japan}  
\author{K. Kitamura} 
\affiliation{Department of Physics, Konan University, Hyogo 658-8501, Japan}
\author{N. Kobayashi} 
\affiliation{Research Center for Nuclear Physics, Osaka University, Osaka 567-0047, Japan}
\author{Y. Matsuda}
\affiliation{Department of Physics, Konan University, Hyogo 658-8501, Japan}
\affiliation{Cyclotron and Radioisotope Center, Tohoku University, Sendai 980-8578, Japan}
\author{A. Nakagawa} 
\affiliation{Cyclotron and Radioisotope Center, Tohoku University, Sendai 980-8578, Japan}
\author{S. Nakamura} 
\affiliation{Research Center for Nuclear Physics, Osaka University, Osaka 567-0047, Japan}
\author{K. Nosaka}
\affiliation{Department of Physics, Konan University, Hyogo 658-8501, Japan} 
\affiliation{Cyclotron and Radioisotope Center, Tohoku University, Sendai 980-8578, Japan}
\author{S. Okamoto}
\affiliation{Department of Physics, Kyoto University, Kyoto 606-8502, Japan}
\author{S. Ota} 
\affiliation{Research Center for Nuclear Physics, Osaka University, Osaka 567-0047, Japan}
\author{S. Pal}
\affiliation{Department of Physics, IIT-ISM Dhanbad, Jharkhand 826004, India}
\author{S. Roy}
\thanks{Present address: IIT Roorkee, India.}
\affiliation{Department of Physics, IIT-ISM Dhanbad, Jharkhand 826004, India}
\author{S. Weyhmiller}
\thanks{Present address: Yale University, USA.}
\affiliation{Department of Physics, University of Notre Dame, Notre Dame, Indiana 46556, USA}
\author{Z. Yang}
\affiliation{Research Center for Nuclear Physics, Osaka University, Osaka 567-0047, Japan}
\affiliation{School of Physics and State Key Laboratory of Nuclear Physics and Technology, Peking University, Beijing 100871, China}
\author{J. C. Zamora}
\affiliation{Facility for Rare Isotope Beams, Michigan State University, East Lansing, Michigan 48824, USA}

\date{\today}
\begin{abstract}
To study the isoscalar giant resonances in a deformed case, background-free $\alpha$-particle inelastic scattering measurements using a 386 MeV $\alpha$ beam were performed on the highly-deformed $^{172}$Yb nucleus using the Grand Raiden spectrometer at the Research Center for Nuclear Physics (RCNP) at very forward angles, including $0^\circ$. The strength distributions for the isoscalar giant resonances up to $L \leq 3$ were obtained using multipole decomposition analysis. The isoscalar giant monopole resonance (ISGMR) strength exhibits a splitting into two components, interpreted as the coupling of the ISGMR with the $K=0$ component of the isoscalar giant quadrupole resonance (ISGQR). A \textit{bimodal} structure is observed in the strength distribution of the isoscalar giant dipole resonance. The ISGQR strength shows an enhancement near 25 MeV, attributed to the excitation of an overtone mode, while the broadening of the main-tone peak is associated with nuclear deformation. The experimental results are well reproduced by theoretical strength distributions calculated using the quasiparticle finite amplitude method for $L \leq 3$.
\end{abstract}
\maketitle


\section{Introduction}
Giant resonances (GRs) are high-frequency, damped, nearly harmonic oscillations of nuclei about their equilibrium shape or density, induced by a weak external probe \cite{Harakeh2001}. These resonant transitions from the ground state to collective states involve changes in spin, isospin, and angular momentum. Among them, the isoscalar giant monopole resonance (ISGMR) or ``breathing mode" and the isoscalar giant dipole resonance (ISGDR) or ``squeezing mode", known as the compressional modes~\cite{Harakeh2001}, which are used to obtain the incompressibility of finite nuclei.

Isoscalar giant resonances (ISGRs) have been well studied in spherical nuclei~\cite{Harakeh2025, Garg_ISGDR1999, Gaidarov2023} and have been studied in several deformed nuclei, such as the Sm isotopes~\cite{Garg1980, Itoh2002, Itoh2003, Youngblood2004}, Nd isotopes~\cite{Garg1984, Abdullah2024}, in the fission decay of $^{238}$U~\cite{Brandenburg1982}, and in lighter-mass nuclei like $^{24}$Mg~\cite{YKGupta2015} and $^{28}$Si~\cite{Peach2016}. Inelastic $\alpha$-particle scattering is commonly used to excite these modes due to the zero spin and zero isospin nature of the $\alpha$ particle. In deformed nuclei, ISGR strength distributions exhibit splitting into different $K$ components~\cite{Garg2018}, where $K$ is the projection of angular momentum onto the symmetry axis. Although the ISGMR in spherical nuclei appears as a single peak, in deformed nuclei it splits into low-energy (LE) and high-energy (HE) components due to coupling with the $K=0$ component of the isoscalar giant quadrupole resonance (ISGQR)~\cite{Garg1980, MBuenerd1980, YAbgrall1980}. For the ISGQR, deformation leads to a splitting into three components $K = 0$, 1, and 2~\cite{Kishimoto1975, YAbgrall1980, Zawischa1978}, which contributes to the broadening of the resonance width~\cite{Kishimoto1975, Yoshida2013}. The ISGMR in $^{172}$Yb has so far been theoretically studied under the quasiparticle random-phase approximation (QRPA) framework using the SV-bas interaction~\cite{JKvasil2016, Klupfel2009}, predicting an LE peak below 10 MeV and an HE peak near 15 MeV. So far, no theoretical studies have been reported for other ISGR modes.

The IVGDR has been extensively studied in both spherical and deformed nuclei through photoneutron cross-section and relativistic Coulomb excitation measurements over a wide mass range~\cite{Berman1975, Plujko2018, Gurevich1981}. The IVGDR exhibits a single Lorentzian peak in spherical nuclei, while in deformed nuclei it exhibits a splitting into two components, corresponding to oscillations along the long and short axes of the prolate shape. The isoscalar giant dipole resonance (ISGDR) has, however, not been studied as much. A \textit{bimodal} strength distribution, consisting of LE and HE components, has been observed in Sm~\cite{Itoh2003} and Nd~\cite{PRC_MA} isotopes. The LE component, associated with toroidal or vortex-like motion~\cite{JKvasil2011, ARepko2013}, exhibits increasing width and energy-weighted sum rule (EWSR) fraction with deformation, while the HE component remains largely unchanged~\cite{Itoh2003}, indicating a transfer of strength from HE to LE~\cite{Itoh2002, Itoh2003}. It should be noted, however, that in Ref.~\cite{HPMorsch1982} the ISGDR was not separated from the high-energy octupole resonance (HEOR), as both resonances appear at similar excitation energies. The ISGDR and HEOR are predicted to couple in deformed nuclei in the $K=0$ and $K=1$ channels, according to QRPA calculations with Skyrme energy-density functionals~\cite{Yoshida2013} and fluid-dynamical models using the generalized scaling approximation~\cite{Nishizaki1985}. The strength of the HEOR is also affected by nuclear deformation~\cite{Itoh2003, HPMorsch1982}.

Except for the ISGMR and ISGDR, which are excited by the second-order term of the transition operator, ISGRs with multipolarity $L \geq 2$ are mainly excited by the first-order term of the transition operator, often referred to as the main-tone mode. For the ISGMR, the first-order term of the transition operator is constant, whereas for the ISGDR, it corresponds to spurious center-of-mass motion. For the higher multipoles, overtone modes arise from higher-order terms of the operator. In the quadrupole case, the overtone (ISGQR2) corresponds to a $4\hbar\omega$ excitation, representing a third type of compressional mode in addition to the ISGMR and ISGDR. The overtone modes of ISGRs with $L \geq 2$ remained elusive for a long time, until a high-energy $L=2$ resonance was identified in $^{208}$Pb at $26.9 \pm 0.7$ MeV with a width of $6.0 \pm 1.3$ MeV through direct proton decay from the ISGDR state~\cite{Hunyadi2003}. This observation was later confirmed through neutron-decay studies in the same nucleus~\cite{Hunyadi2007}. The overtone of the ISGQR was recently observed for the first time via multipole decomposition analysis (MDA) in Nd isotopes~\cite{PRC_MA}. Continuum-RPA calculations within semi-microscopic frameworks further predict quadrupole overtone modes with centroid energies above 30 MeV~\cite{Gorelik2004}. 

In this work, the ISGRs in the highly deformed nucleus $^{172}$Yb [$\beta_2 = 0.330 (5)$]~\cite{nndc} are investigated to study the effect of ground-state deformation on the strength distributions. Similar studies have been carried out for $^{154}$Sm, which has a comparable deformation [$\beta_2 = 0.339 (3)$]~\cite{nndc}, but the optical model parameters (OMPs) were extracted using the spherical isotope $^{144}$Sm~\cite{Itoh2003, Youngblood2004}. In contrast, the present analysis derives the OMPs directly for the deformed nucleus $^{172}$Yb.

This work presents a comprehensive and systematic investigation of isoscalar giant resonances with $L \leq 3$ in the highly-deformed nucleus $^{172}$Yb through MDA. Notably, an overtone mode is observed in the ISGQR strength distribution at a high excitation energy of approximately 25 MeV. In addition, theoretical strength distributions for isoscalar giant resonances have been calculated using the quasiparticle finite amplitude method (QFAM) and compared with the corresponding experimental data.

\section{Experimental Setup}
\label{ExpS_Disc}
The experiment was performed at the Research Center for Nuclear Physics (RCNP), Osaka University, using the high-resolution Grand Raiden (GR) spectrometer~\cite{Fujiwara1999} at very forward angles. An $\alpha$-particle beam was accelerated to 386 MeV using the AVF and ring cyclotrons, with single-turn extraction from the ring cyclotron providing a high-quality, halo-free $\alpha$ beam~\cite{Fujiwara1999}. At this energy, contributions from multistep processes such as pick-up or knock-out reactions are negligible, resulting in background-free spectra~\cite{Tamii2009}. The beam was delivered through the west-south beamline without the use of slits, further minimizing the beam halo before reaching the target chamber. Due to the zero spin and isospin characteristics of the $\alpha$ particle, it is particularly well suited for investigating isoscalar giant resonance modes. The $\alpha$ beam bombarded a thin, self-supporting $^{172}$Yb target with an areal density of 2.95 $\pm$ 0.08 mg/cm$^2$. The target thickness was determined as the average of the nominal value and the areal density derived from weighing. The associated uncertainty corresponds to the difference between these two measurements. The beam current ranged from 0.1 to 10 nA, constrained by both the data-acquisition system and the performance of the accelerator. The energy resolution of the spectrometer was approximately 175 keV~\cite{Abdullah2024}.

Inelastically scattered $\alpha$ particles were momentum analyzed using the GR spectrometer and directed onto a focal-plane detector system comprising two position-sensitive Multi-Wire Drift Chambers (MWDCs) and two plastic scintillators. The MWDCs provided measurements of the vertical ($Y_{\rm fp}$) and horizontal ($X_{\rm fp}$) positions, while the plastic scintillators measured the energy deposition. This setup allowed particle identification and precise reconstruction of particle trajectories. The position and scattering angle were determined using the ray-tracing method~\cite{patel_thesis2016}. The scattering angles were averaged over the acceptance of the GR spectrometer. At the exit of the GR scattering chamber, collimators defined the vertical acceptances of $\Delta\phi = 40$ mrad at 0$^\circ$ and 60 mrad at other finite-angle settings of the GR spectrometer. The horizontal acceptance was determined later through offline analysis.

Elastic scattering was measured from 3.5$^{\circ}$ to 20.5$^{\circ}$ to extract OMPs. Inelastic-scattering data were taken at forward angles $0^{\circ} \le \theta_{\mathrm{Lab}} \le 10^{\circ}$. To calibrate the excitation-energy spectra for $^{172}$Yb, $\alpha$-particle inelastic scattering on a $^{24}$Mg target with an areal density of 2.5 mg/cm$^2$ was measured under identical spectrometer settings and scattering angles as those used for $^{172}$Yb. The focal-plane detectors were calibrated using the well-characterized low-lying discrete states of $^{24}$Mg identified from high-resolution $^{24}$Mg($\alpha, \alpha'$) data~\cite{Kawabata}. The calibration was done following the procedure in Refs.~\cite{PRC_MA, Abdullah_thesis}. A sieve slit with 5 mm horizontal and 12 mm vertical-hole spacing was placed between the target and the spectrometer magnet for ion-optical corrections. 
\begin{figure}[hbt!]
    \centering
    \includegraphics[width=1.0\linewidth]{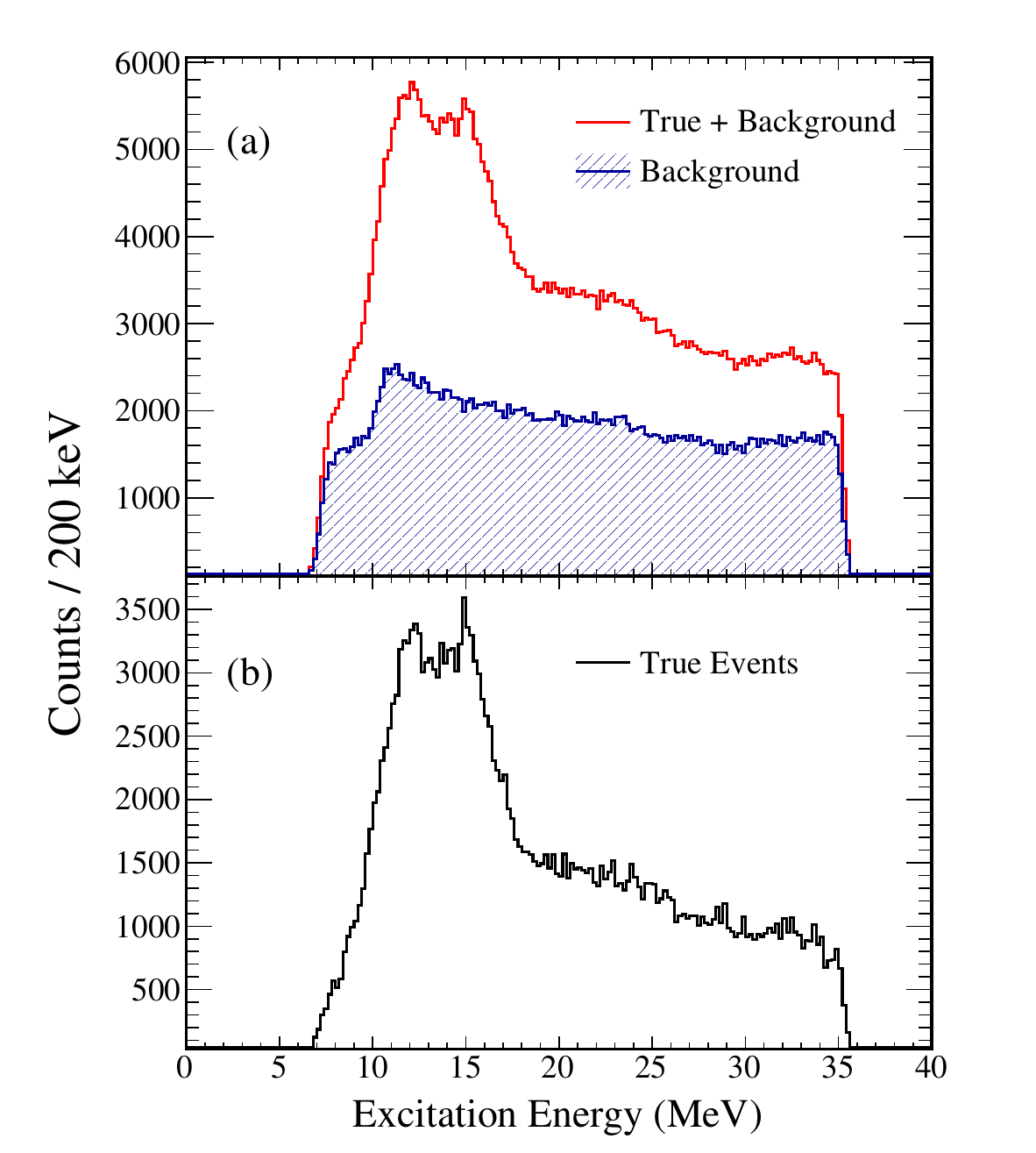}
    \caption{(a) Excitation-energy spectrum of $^{172}$Yb($\alpha, \alpha'$) measured at an average spectrometer angle of $\theta_{\mathrm{avg}}$ = 0.75$^{\circ}$, prior to instrumental background subtraction. The red histogram represents the total (true + background) events, while the blue-hatched area indicates the instrumental background. (b) Excitation-energy spectrum after instrumental background subtraction, where the black curve corresponds to the true inelastic scattering events.}
    \label{bg_subtraction}
\end{figure}

The GR spectrometer was operated in double-focusing mode, enabling instrumental background subtraction. After particle identification, ion-optical correction, and calibration, the excitation-energy spectrum was obtained, as shown in Fig.~\ref{bg_subtraction}, following the procedure of Refs.~\cite{PRC_MA, Abdullah_thesis}. Figure~\ref{bg_subtraction}(a) illustrates the excitation-energy spectrum at the $0^\circ$ GR angle before background subtraction, which corresponds to an average spectrometer angle of $\theta_{\rm avg} = 0.75^{\circ}$. The blue-hatched area indicates the instrumental background. This background remains essentially flat across the entire excitation-energy range. It reaches approximately 40$\%$ of the total spectrum at 0$^\circ$, but it drops rapidly with increasing spectrometer angle. Figure~\ref{bg_subtraction}(b) shows the excitation-energy spectrum after background subtraction. As $^{172}$Yb is highly deformed, the excitation-energy spectrum exhibits splitting, as shown in Fig.~\ref{bg_subtraction}.

\section{Data Analysis}

\subsection{Optical Model Parameters extraction from Elastic-Scattering data}
\label{OMP_Disc}
To investigate the angular distributions from inelastic-scattering data, calculated within the framework of the distorted-wave Born approximation (DWBA), the OMPs are essential parameters. These OMPs are extracted from elastic-scattering data. For this purpose, $\alpha$-particle elastic scattering on $^{172}$Yb was performed over a broad GR angular range from 3.5$^{\circ}$ to 20.5$^{\circ}$.

\begin{figure}[hbt!]
    \includegraphics[width=1.0\linewidth]{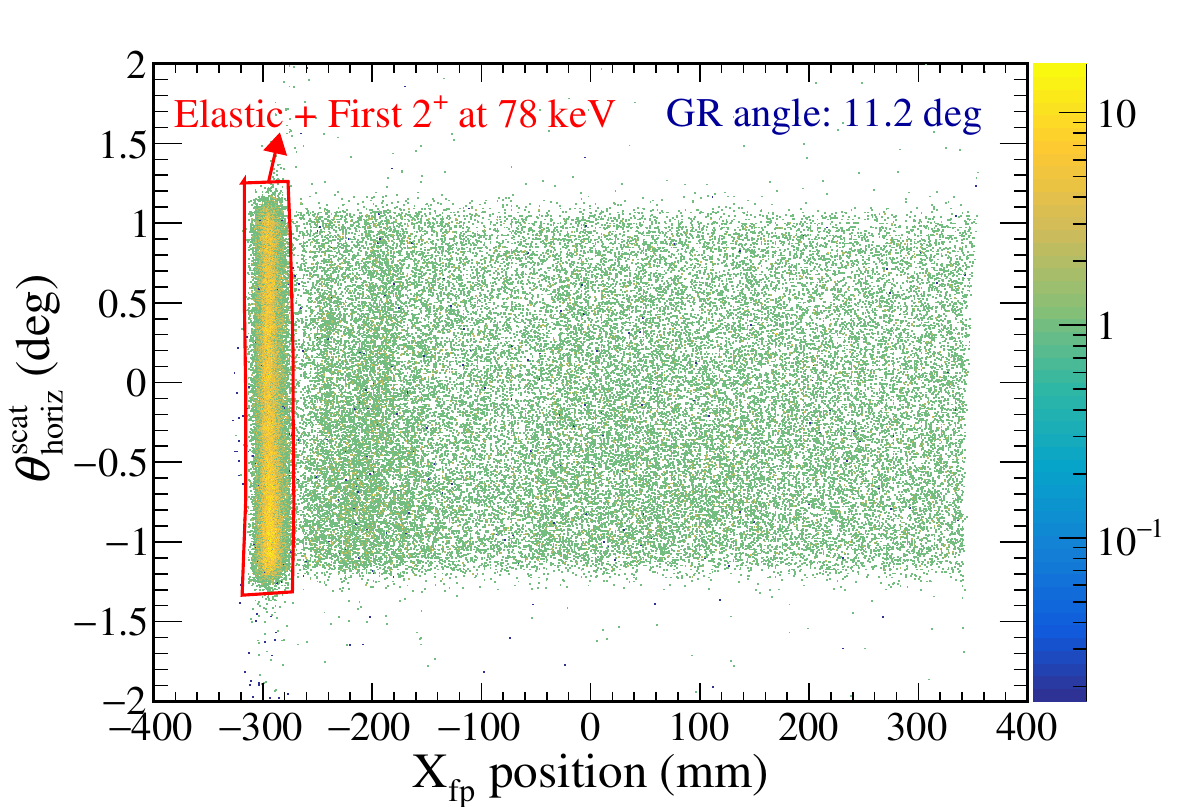}
    \caption{Two-dimensional spectrum of $^{172}$Yb($\alpha,\alpha$) elastic scattering at the GR angle of 11.2$^{\circ}$ without GR averaging. The red g-cut encloses the overlapping elastic peak and first-excited 2$^{+}$ state at 78 keV.}
    \label{theta_xfp}
\end{figure}

\begin{figure}[hbt!]
    \includegraphics[width=1.0\linewidth]{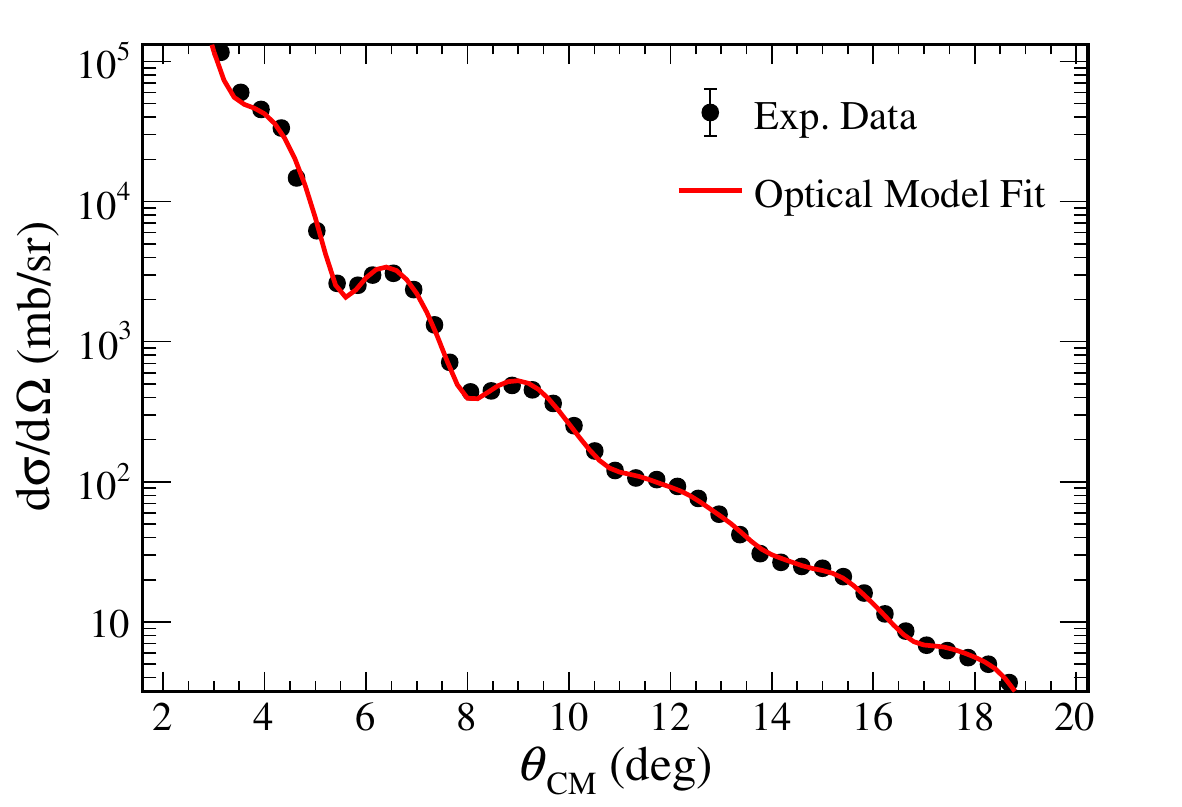}
    \caption{Fitting of the EPFE2 peak for $^{172}$Yb($\alpha,\alpha$). The error bars are within the data points. The black filled data points are the measured experimental cross sections, while the red solid line shows the optical-model fit.}
    \label{elastic_fit}
\end{figure}

After particle identification, background subtraction, and ion-optical correction, a two-dimensional histogram of the elastic-scattering data was generated following the procedure in Refs.~\cite{PRC_MA, Abdullah_thesis}, as shown in Fig.~\ref{theta_xfp}. The spectrum is presented without GR averaging. As mentioned earlier, the energy resolution of the spectrometer during the experiment was 175 keV. Due to this poor resolution, the elastic scattering peak and the 2$^{+}$ inelastic scattering peak (EPFE2) at 78 keV were merged, as shown in Fig.~\ref{theta_xfp}. The higher-lying excited states beyond the EPFE2 peak were also overlapping with other states and merged into the background, making them indistinguishable, as shown in Fig.~\ref{theta_xfp}. To evaluate the cross section of the EPFE2 peak, a red two-dimensional contour (g-cut) was applied around the corresponding region. Within this g-cut, the $\theta^{\text{scat}}_{\text{horiz}}$ was divided into four equal intervals, each of width 0.4$^{\circ}$, spanning the range from $-$0.8$^{\circ}$ to +0.8$^{\circ}$. The cross section was extracted for each angular bin for the EPFE2 peak, and the corresponding $\theta^{\text{scat}}_{\text{horiz}}$ values were averaged over the GR angle. This procedure was repeated for each GR angle. The average scattering angles, $\theta_{\text{avg}}$, were converted to center-of-mass angles, $\theta_{\text{CM}}$. The corresponding differential cross sections were then plotted as a function of $\theta_{\text{CM}}$, representing the angular distribution of the EPFE2 peak (see Fig.~\ref{elastic_fit}). A detailed discussion will be provided in Ref.~\cite{Kalpana_thesis}.

The optical model and DWBA calculations were performed using the coupled-channel code CHUCK3~\cite{Chuck3}, employing a complex potential of the form $V(r) + iW(r)$, where the real $V(r)$ and the imaginary $W(r)$ terms are of the Woods-Saxon type. For the real potential this is given by: 
\begin{equation}
 V(r) = -{\frac{V{_R}} {\left[1 + \exp\left(\frac{r-R{_R}}{a{_R}}\right)\right]}},
 \label{Potential_form}
\end{equation}
where $V_{R}$, $R_{R}$, and $a_{R}$ are the depth, radius, and diffuseness for the real potential, respectively. While fitting, the value of the Coulomb radius $R_c$ was fixed at 1.30 fm, consistent with Ref.~\cite{Morsch1982} for actinide nuclei.

To determine the optimal set of OMPs, $\chi^2$ minimization was performed using both brute-force and basin-hopping optimization techniques implemented in a Python framework \cite{Das2021}. The theoretically calculated angular distributions for both the elastic scattering and the first-excited 2$^{+}$ state, obtained using CHUCK3, were also added during the $\chi^2$ minimization. An initial arbitrary set of six OMPs was supplied to CHUCK3, which generated corresponding angular distributions. These theoretical distributions were compared to the experimental data, and the $\chi^2$ values were evaluated. The optimization algorithms iteratively adjusted the OMPs, feeding each new set into CHUCK3 to compute updated angular distributions, thereby refining the fit. The process continued until convergence to a minimum $\chi^2$ value was achieved. The final fit to the experimental cross-section data is presented in red in Fig.~\ref{elastic_fit}, while the optimized set of OMPs is summarized in Table~\ref {OMPs_table}.

\begin{table}[htb!]
\centering
\begin{tabular}{cccccccc}
\hline
\hline
$E_{\alpha}$ & $V_R$  & $V_I$ & $R_R$ & $R_I$ & $a_R$ & $a_I$ & $R_c$ \\
(MeV) & (MeV) & (MeV) & (fm) & (fm) & (fm) & (fm) & (fm)
\\
\hline
\\
386 & 163.3 & 25.49 & 5.61 & 8.12 & 1.060 & 0.609 & 1.30 \\
\hline
\hline
\end{tabular}
\caption{OMPs obtained from the fitting of
the EPFE2 peak after $\chi^2$ minimization.}
\label{OMPs_table}
\end{table}


\subsection{Multipole Decomposition Analysis}
\label{MDA_Disc}
\begin{figure*}[hbt!]
    \centering
    \includegraphics[width=0.85\linewidth]{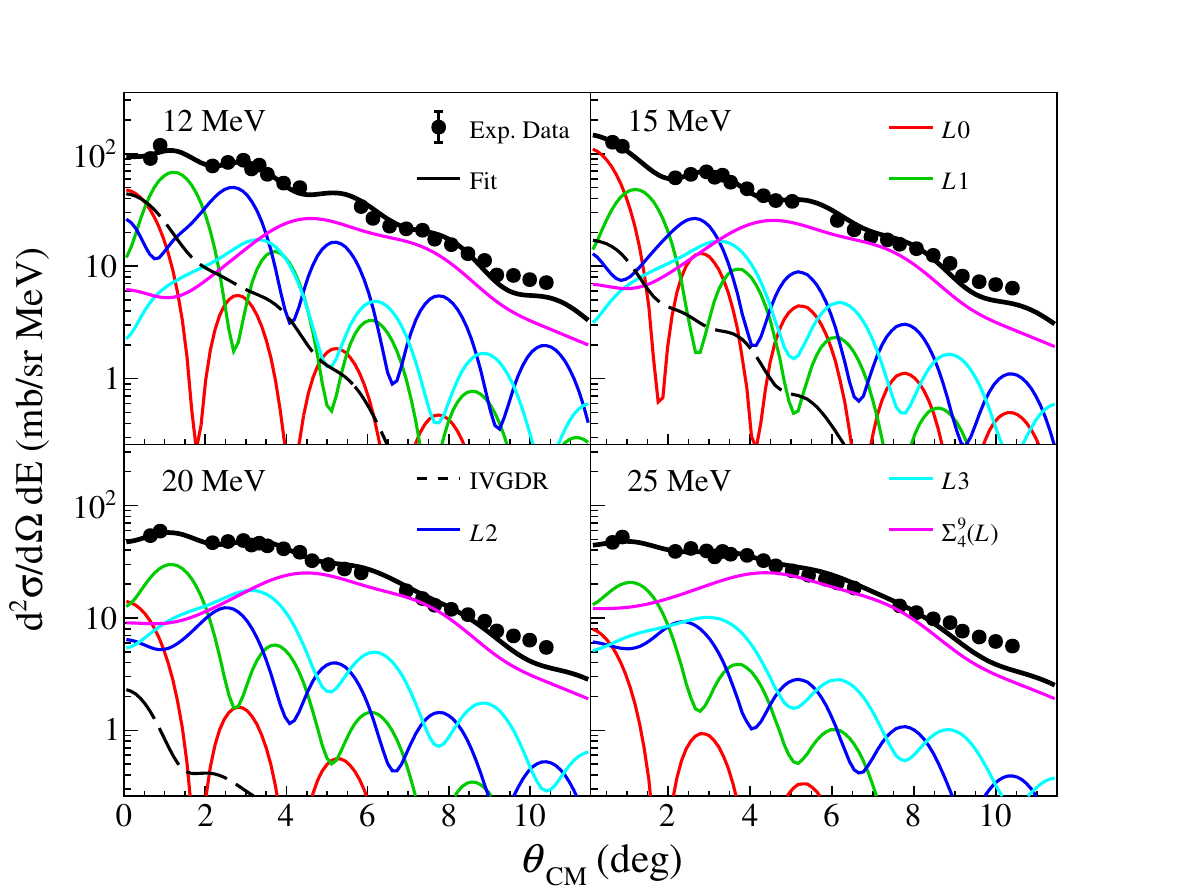}
    \caption{Double-differential cross sections at 12, 15, 20, and 25 MeV for $^{172}$Yb are presented. The solid black curves represent fits from the MDA. Contributions from individual multipoles are shown: $L = 0$ (red), $L = 1$ (green), $L = 2$ (blue), $L = 3$ (cyan), $L\geq4$ (magenta), along with the IVGDR component (dashed black lines). The error bars are within the data points. The missing data points for angles between 4$^{\circ}$ and 8$^{\circ}$ in the different plots correspond to the regions where the hydrogen contamination was removed.}
    \label{MDA_fit}
\end{figure*}

The background-free, calibrated excitation-energy spectra of $^{172}$Yb were used to determine the cross section, which contains contributions from various multipoles. To disentangle these multipoles, an MDA was performed. The excitation-energy spectra were binned in 1 MeV intervals. This binning choice was made to enhance statistics. Within each 1 MeV excitation-energy bin, the scattering angle $\theta^{\text{scat}}_{\text{horiz}}$ was further partitioned into four equal segments of 0.4$^\circ$ each, covering the range from $-0.8^\circ$ to $0.8^\circ$ and then averaged. The differential cross sections corresponding to these angular selections were then determined in each energy bin following the procedure outlined in Sec.~\ref{OMP_Disc}. The differential cross-section data are shown in Fig.~\ref{MDA_fit}, with error bars within the data points. The errors in the data points represent both statistical and systematic uncertainties, where the systematic error arises from the uncertainty in target thickness, as described in Sec.~\ref{ExpS_Disc}. The extracted OMPs from the EPFE2 peak data were used for DWBA calculations to obtain the angular distribution. These calculations were done with the CHUCK3 code~\cite{Chuck3} for a wide excitation-energy range using the transition densities and 100$\%$ sum rules for various multipolarities. The transition densities~\cite{Harakeh2001}, sum rules, and DWBA calculations are detailed in Refs.~\cite{PRC_MA, Abdullah_thesis}. Contributions from hydrogen contamination to the spectra at several forward scattering angles were identified as broad bumps overlapping with the giant-resonance bumps. These were removed from the angular distributions for the excitation energy bins of $^{172}$Yb for which this overlap occurred. Before performing MDA, the IVGDR strength contribution, arising from Coulomb excitation, had to be subtracted from the excitation-energy spectra. The IVGDR strength for $^{172}$Yb was taken from Ref.~\cite{Kleinig2008}, where it was calculated using the Skyrme random-phase approximation and used to estimate the exhausted sum rules. The IVGDR form factors were obtained from the Goldhaber–Teller model~\cite{Satchler1987}. 

In MDA, the experimentally obtained differential cross section is expressed as a linear combination of calculated angular distributions for various multipoles within the DWBA framework.
\begin{equation}
\frac{d^2\sigma^{\rm exp}(\theta_{\rm CM}, E_{\rm x})}{d\Omega dE} = \sum_{L = 0}^{9} A_{L}(E_{\rm x}) \frac{d^2\sigma_L^{\rm DWBA}(\theta_{\rm CM}, E_{\rm x})}{d\Omega dE},
\label{MDA_Eqn}
\end{equation}

where $\frac{d^2\sigma^{\rm exp}(\theta_{\rm CM}, E_{\rm x})}{d\Omega dE}$ denotes the experimental angular distribution after subtracting the instrumental background, IVGDR, and hydrogen contamination. The $\frac{d^2\sigma_L^{\rm DWBA}(\theta_{\rm CM}, E_{\rm x})}{d\Omega dE}$ term is the calculated angular distribution of the $L${th} multipole in the DWBA framework corresponding to 100$\%$ energy-weighted sum rule (EWSR). The $A_{L}(E_{\rm x})$ is the fraction of the EWSR for the $L${th} multipole in each energy bin, calculated by minimizing $\chi{^2}$.\\
In the MDA fit, multipolarities up to $L = 9$ are included, but the strength is extracted only for $L\leq3$. The angular distributions for $L\geq4$ are summed up and included as a single component in the fit. The MDA fits for excitation energies 12, 15, 20, and 25 MeV are shown in Fig.~\ref{MDA_fit}. The poor fit of the last three data points may be due to the calculation of the OMPs from the EPFE2 peak data. The fractions of the EWSR $A_{L}(E_{\rm x})$, obtained from the MDA, were used to calculate the strength distributions for each multipole. The strength distributions were obtained using equations from Ref.~\cite{PRC_MA}.


\section{Theoretical Framework: QFAM Calculation}

Theoretical strength distributions for the isoscalar giant resonances ($L\leq3$) in $^{172}$Yb 
were calculated using the QFAM approach~\cite{ABjelcic_2020_CPC_DIRQFAM,ABjelcic_2023_CPC_DIRQFAM2}. 
The QFAM approach provides an efficient way for computing the linear response in superfluid nuclei, 
particularly useful for deformed systems where traditional QRPA calculations can be computationally intensive.
The calculations were performed on top of the ground state obtained from the axially deformed relativistic Hartree-Bogoliubov (RHB) model~\cite{TNiksic_2014_CPC_DIRHB}.
The RHB Hamiltonian $H$ is derived from the variation of an energy density functional $E[R]$ 
with respect to the generalized density matrix $R$, which consists of both the density matrix and the pairing tensor.

The starting point of the QFAM approach is the time-dependent RHB model.
The time-dependent RHB equation,
\begin{equation}\label{eq:tdfhb}
    \mathrm{i}\hbar \dot{R}(t) = [H\left[R(t)\right]+F(t),R(t)],
\end{equation}
describes the nuclear response to an external field,
\begin{equation}
    \hat{F}(t)
    = \eta \left\{ \hat{F}(\omega) \mathrm{exp}(-\mathrm{i} \omega t)
    + \hat{F}^{\dagger} (\omega) \mathrm{exp}(\mathrm{i} \omega t) \right\}.
\end{equation}
Here, the linear response of the nucleus is considered, so that a weak external field is adopted, where $\eta$ is taken as an infinitesimal real parameter and $\omega$ denotes the oscillation frequency.
For the ISGMR, ISGDR, ISGQR, and isoscalar giant octupole resonance (ISGOR), the external fields are defined as,
\begin{equation}
    \begin{aligned}\label{QFAM_Op}
        \hat{F}^{\mathrm{ISGMR}}   = {}& \sum_k r_k^2,\\
        \hat{F}^{\mathrm{ISGDR}}_K = {}& \sum_k (r_k^3 - \zeta r_k) Y_{1K}(\hat{r}_k),\\
        \hat{F}^{\mathrm{ISGQR}}_K = {}& \sum_k r_k^2 Y_{2K}(\hat{r}_k),\\
        \hat{F}^{\mathrm{ISGOR}}_K = {}& \sum_k r_k^3 Y_{3K}(\hat{r}_k),
    \end{aligned}
\end{equation}
where $k$ runs over all nucleons and $\zeta$ is a parameter used to eliminate the spurious center-of-mass motion in the ISGDR~\cite{ARepko_2019_PRC_spurious}.

The induced change in the generalized density matrix can be decomposed into components with specific frequencies,
\begin{equation}
    \delta R(t)= \eta\left\{\delta R(\omega) \mathrm{exp}(-\mathrm{i} \omega t)+\delta R^{\dagger} (\omega) \mathrm{exp}(\mathrm{i} \omega t)\right\}.
\end{equation}
Similarly, the RHB Hamiltonian admits an expansion of the form,
\begin{equation}
    H(t)=H_{0}+\eta\left\{\delta H(\omega) \mathrm{exp}(-\mathrm{i} \omega t)+\delta H^{\dagger} (\omega) \mathrm{exp}(\mathrm{i} \omega t)\right\}.
\end{equation}

By linearizing the equation of motion Eq.~\eqref{eq:tdfhb} with respect to the small amplitude parameter $\eta$
and expanding all operators in the quasiparticle basis, one obtains the linear-response equations in the frequency domain,
\begin{equation}
    \begin{aligned}
        \left(E_{\mu}+E_{\nu}-\omega\right) X_{\mu \nu}(\omega)+\delta H^{20}_{\mu \nu}(\omega)= & -F^{20}_{\mu \nu}(\omega), \\
        \left(E_{\mu}+E_{\nu}+\omega\right) Y_{\mu \nu}(\omega)+\delta H^{02}_{\mu \nu}(\omega)= & -F^{02}_{\mu \nu}(\omega).
    \end{aligned}
\end{equation}
Here, the matrix elements $F^{20}(\omega)$ [$\delta H^{20}(\omega)$] and $F^{02}(\omega)$ [$\delta H^{02}(\omega)$] 
correspond to the two-quasiparticle creation and annihilation components of the external field (induced RHB Hamiltonian), 
respectively, expressed in the quasiparticle basis.
The deviation $\delta R(\omega)$ of the generalized density matrix from the ground state 
is defined via the amplitudes $X(\omega)$ and $Y(\omega)$ as,
\begin{equation}
    \delta R(\omega)=\left(\begin{matrix} 0 & \delta R^{20}(\omega) \\ - \delta R^{02}(\omega) & 0 \end{matrix}\right)
    \equiv\left(\begin{matrix} 0 & X(\omega) \\ - Y(\omega) & 0 \end{matrix}\right).
\end{equation}
In practice, these equations are solved through an iterative procedure that requires only the first derivatives of $E[R]$ with respect to $R$.
This strategy circumvents the need for matrix diagonalization and the computationally intensive evaluation of two-body matrix elements in the conventional QRPA.
A comprehensive description of the QFAM formalism can be found in Refs.~\cite{TNiksic_2013_PRC_QFAM,ABjelcic_2020_CPC_DIRQFAM,ABjelcic_2023_CPC_DIRQFAM2}.
In the present work, the same nuclear Hamiltonian is used consistently in the RHB and QFAM calculations.
The ph channel of the effective interaction is derived from the relativistic point-coupling density functional DD-PC1~\cite{TNiksic_2008_PRC_DDPC1}.
For the pp channel, a separable pairing force is adopted, as introduced in Refs.~\cite{YTian_2009_PLB_pairing, YTian_2009_PRC_axialpairing, YTian_2009_PRC_pairing}.

The strength function can be obtained by,
\begin{equation}
    S(\omega) = -\frac{1}{\pi} \operatorname{Im}\left[ \frac{1}{2} \sum_{\mu\nu} F^{20\ast}_{\mu\nu} X_{\mu\nu}(\omega) + F^{02\ast}_{\mu\nu} Y_{\mu\nu}(\omega)\right], 
\end{equation}
where a complex frequency $ \omega\rightarrow\omega+\mathrm{i}\gamma$ is introduced, and
a Lorentzian smearing with a width of $2\gamma$ is obtained.
Giant resonances exhibit broad structures whose widths originate from damping mechanisms related to collective motions. These properties are not fully captured within the standard QRPA or QFAM frameworks.
A more comprehensive treatment can be achieved by incorporating higher-order many-body correlations, 
such as two-particle-two-hole configurations~\cite{Gambacurta_2020_PRL,CLBai_2010_PRL} 
or (quasi-) particle vibration coupling effects~\cite{YFNiu_2015_PRL,YFNiu_2018_PLB,ZZLi_2023_PRL,ZZLi_2024_PRC,Litvinova_2019_PRC,YNZhang_2022_PRC}. 
However, such extensions entail significantly higher computational costs, especially in deformed nuclei.
In this work, a phenomenological description of damping is adopted by adjusting $\gamma$ to match the experimental data.
Consequently, $\gamma$ is set to 1 MeV, 0.5 MeV, 1.5 MeV, and 0.5 MeV for the ISGMR, ISGDR, ISGQR, and ISGOR, respectively.

The RHB and QFAM equations are solved numerically by expanding the wave functions in the axially deformed harmonic oscillator basis and the simplex-y HO basis, respectively. A truncation with a maximum principal quantum number of $20$ is applied to ensure numerical convergence. The self-consistent calculation predicts a prolate deformation ($\beta_2=0.337$) for the ground state of $^{172}$Yb, with a binding energy of $1392.485$ MeV. These results are in good agreement with the experimental values of $E=1392.758$ MeV and $\beta_2=0.330$.

\section{Results and Discussion}


\subsection{ISGMR}
\label{ISGMR_Disc}
\begin{figure}[hbt!]
    \centering
 \includegraphics[width=1.1\linewidth]{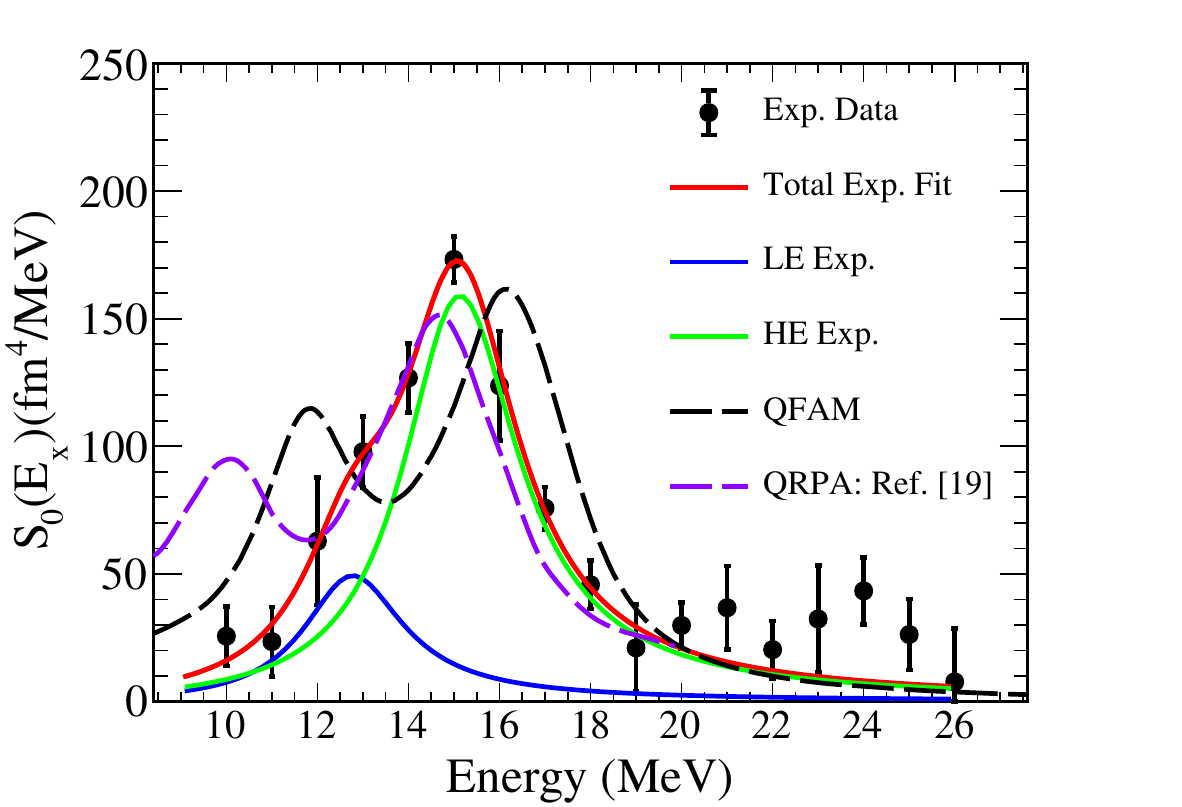}
    \caption{ISGMR strength distribution in $^{172}$Yb extracted from MDA. The error bars on the data points include systematic uncertainties. The thick solid red line shows the double Lorentzian fit to the experimental data, with LE and HE components indicated by thick solid blue and green lines, respectively. The black dashed line represents the QFAM theoretical prediction, while the purple dashed line corresponds to the QRPA strength with the SV-bas interaction~\cite{JKvasil2016}.}
    \label{ISGMR}
\end{figure}

The transition operator employed to extract the isoscalar monopole strength is
\begin{equation} 
O^{00} = \frac{Z}{2A} \sum_k r_k^{2},
\label{Operator_Monopole} 
\end{equation}

where the summation runs over all nucleons. The factor $Z/A$ accounts for fully collective isoscalar transitions, in which all nucleons contribute coherently to the collective motion~\cite{Harakeh2001}. This represents a second-order (overtone) term in the expansion of the spherical Bessel function, as the first-order term is a constant and does not contribute to intrinsic nuclear excitation. The ISGMR ($L=0$) strength distribution in $^{172}$Yb was extracted through MDA using the equation given in Ref.~\cite{PRC_MA}. The ISGMR strength distribution is shown in Fig.~\ref{ISGMR}. The uncertainties in the experimental data points include both systematic errors and the errors arising from the MDA fitting procedure. The extracted strength showed a dependence on the specific choice of $\theta^{\text{scat}}_{\text{horiz}}$ selection at the nominal $0^\circ$ GR angle. To quantify the systematic uncertainty, different angular bin widths were varied to examine their effect on the extracted strength distribution. The resulting fluctuations in the strength distribution were properly taken into account as systematic uncertainties. The $^{172}$Yb nucleus is characterized by prolate axial deformation with a quadrupole deformation parameter $\beta{_2} = 0.33 (5)$~\cite{nndc}. This deformation induces coupling of the ISGMR with the $K=0$ component of the ISGQR, leading to a splitting of the ISGMR strength into LE and HE components.

\begin{table*}[hbt!]
\centering
\caption{Experimental fitting parameters of the ISGMR strength in $^{172}$Yb, including the centroid energy $E_{\rm x}$, width $\Gamma$, and EWSR for both the LE and HE components. The quoted uncertainties on $E_{\rm x}$, $\Gamma$, and EWSR correspond to the 68\% confidence interval.}
\renewcommand{\arraystretch}{2.0}
\setlength{\tabcolsep}{21.9pt}
\begin{tabular}{lcccccc}
\hline
\hline
 \multicolumn{3}{c}{LE component} & \multicolumn{3}{c}{HE component} \\
 $E_{\rm x}$ (MeV) & $\Gamma$ (MeV) & EWSR (\%) & $E_{\rm x}$ (MeV) & $\Gamma$ (MeV) & EWSR (\%) \\
\hline
 $12.8 \pm 0.4$ & $2.7 \pm 0.7$ & $15.6^{+8.5}_{-6.6}$ & $15.1 \pm 0.1$ & $3.1 \pm 0.2$ & $67.6^{+8.0}_{-7.6}$ \\
\hline
\hline
\label{ISGMR_Tab}
\end{tabular}
\end{table*}

The experimental data of the ISGMR strength distribution were fitted with a double Lorentzian function in the excitation-energy range of 10 to 19 MeV, as shown by the solid red line in Fig~\ref{ISGMR}. The functional form of the fit is described in Ref.~\cite{PRC_MA}, and the extracted fitting parameters are summarized in Table~\ref{ISGMR_Tab}. The Lorentzian fit to the ISGMR strength distribution was performed using a $\chi^2$-minimization procedure with all parameters treated as free variables. Parameter uncertainties were evaluated by varying one parameter while keeping the others fixed at their best-fit values. In this manner, all parameter uncertainties were constrained within the 68$\%$ confidence interval, and the uncertainty in the extracted EWSR was obtained through propagation of the errors of the extracted fitting parameters. The EWSR for the LE (solid blue line) and HE (solid green line) components in Fig.~\ref{ISGMR} was obtained by integrating $E_{\rm x} S_0(E_{\rm x})$ in the excitation-energy range of 10 to 26 MeV, where $S_{0}(E_{\rm x})$ was estimated from the Lorentzian fits. The $\%$EWSR values of the LE (HE) peak in $^{172}$Yb are comparable to those reported for $^{154}$Sm~\cite{Itoh2003}. The enhancement of ISGMR strength above 20 MeV as seen in Fig.~\ref{ISGMR} arises from knock-out and quasi-free processes~\cite{Garg2018}.

Figure~\ref{ISGMR} also shows the experimental data compared with the theoretical strength distributions obtained within the QFAM framework (black dashed line) and QRPA calculation using the SV-bas interaction (purple dashed line)~\cite{JKvasil2016}. In the QRPA calculation, the monopole transition operator is taken as $\sum_{i=1}^{A} (r^{2} Y_{00})_{i}$~\cite{JKvasil2016}. This operator differs from that in Eq.~\ref{Operator_Monopole} by a factor of $Z\sqrt{\pi}/A$. Consequently, the QRPA strengths from Ref.~\cite{JKvasil2016} were scaled by $Z^{2}\pi/A^{2}$ to ensure consistency between the operator definitions. A difference of $Z/2A$ exists between the monopole transition operator employed in the present experimental analysis (see Eq.~\ref{Operator_Monopole}) and the operator adopted in the QFAM calculations for the ISGMR (see Eq.~\ref{QFAM_Op}). By scaling the QFAM strength distribution with a factor of $Z^2/4A^2$, the resulting expression becomes identical to the operator form of Eq.~\ref{Operator_Monopole}, in agreement with the strength formulas reported in Ref.~\cite{PRC_MA}. The experimental strength distribution is in overall agreement with the QFAM results. The centroid energy and width of the LE (HE) peak in the experimental results are consistent with those of the corresponding peak obtained from QFAM calculation within the 1 MeV energy bin. The QRPA calculation reproduces the HE peak well but predicts the LE peak at an excitation energy below 10 MeV, significantly lower than observed experimentally.

\begin{figure}[hbt!]
    \centering
 \includegraphics[width=1.00\linewidth]{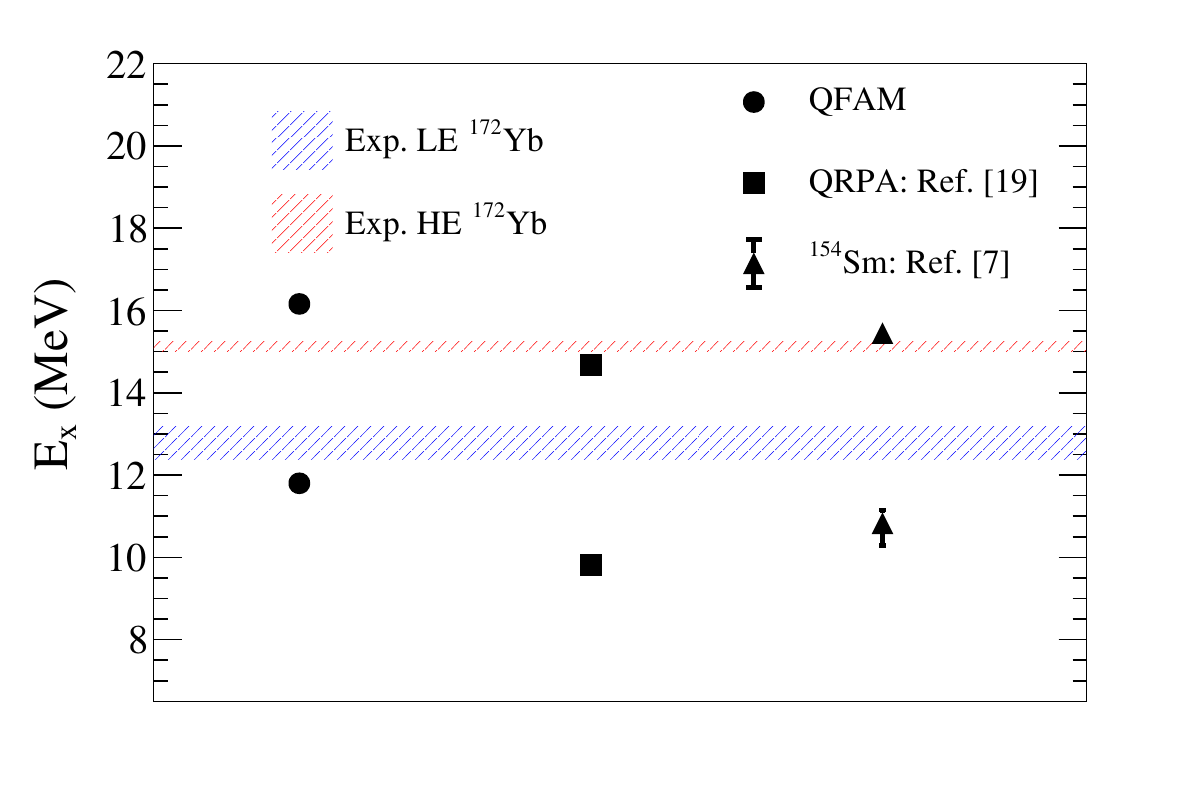}
\caption{Centroid energies of the LE and HE peaks. The experimental LE centroid energy in $^{172}$Yb is shown as a blue horizontal hatched band, with the band width indicating the associated error. The HE centroid energy in $^{172}$Yb is shown as a red horizontal hatched band, with the band width representing its error. The QFAM centroid energies for LE and HE in $^{172}$Yb are shown as black filled circles, while those from QRPA calculations with the SV-bas interaction~\cite{JKvasil2016} are shown as black filled squares. For comparison, the experimental centroid energies for LE and HE in $^{154}$Sm~\cite{Itoh2003} are shown as black filled triangles.}
    \label{centroid_energy}
\end{figure}

This is further seen in Fig.~\ref{centroid_energy}, which shows the experimentally extracted centroid energies of the LE and HE peaks in $^{172}$Yb compared with the QFAM results, QRPA calculations~\cite{JKvasil2016}, and experimental data for $^{154}$Sm~\cite{Itoh2003}. The experimental LE peak in $^{172}$Yb is represented by a blue horizontal hatched band, and the HE peak by a red horizontal hatched band, with band widths indicating the respective uncertainties. The HE centroid energies from QFAM, QRPA, and $^{154}$Sm agree with the experimental value for $^{172}$Yb within a 1 MeV bin, while the LE centroids are consistent within the same uncertainty, except for the QRPA prediction, which lies below 10 MeV, which is outside the acceptance of the spectrometer.

The ratio of the EWSRs of the LE and HE peaks of the experimental ISGMR strength in $^{172}$Yb is found to be $4.3 \pm 1.7$. A comparable value of $4.1 \pm 1.2$ was previously reported for $^{154}$Sm in Ref.~\cite{Itoh2003}, based on measurements performed at RCNP with a 386 MeV $\alpha$ beam. The agreement between these results is reasonable, since $^{154}$Sm has a deformation parameter of $\beta{_2} = 0.339 (3)$~\cite{nndc}, which is similar to that of $^{172}$Yb with $\beta{_2} = 0.33 (5)$~\cite{nndc}. The theoretical calculation with the SkP-based model estimates the ratio to be around $3.2$ for $^{154}$Sm~\cite{Yoshida2013}. In contrast, a lower value of $2.5 \pm 0.2$ was measured for $^{154}$Sm in $\alpha$-particle inelastic scattering at 240 MeV at Texas A$\&$M University~\cite{Youngblood2004}. This reduced ratio indicates enhanced ISGMR-ISGQR mixing. These discrepancies may arise from differences in $\alpha$ beam energies~\cite{JKvasil2016}. Similar systematics have also been reported for Nd isotopes~\cite{PRC_MA}.


\subsection{ISGDR}
\label{ISGDR_Disc}
\begin{figure}[hbt!]
    \centering
    \includegraphics[width=1.1\linewidth]{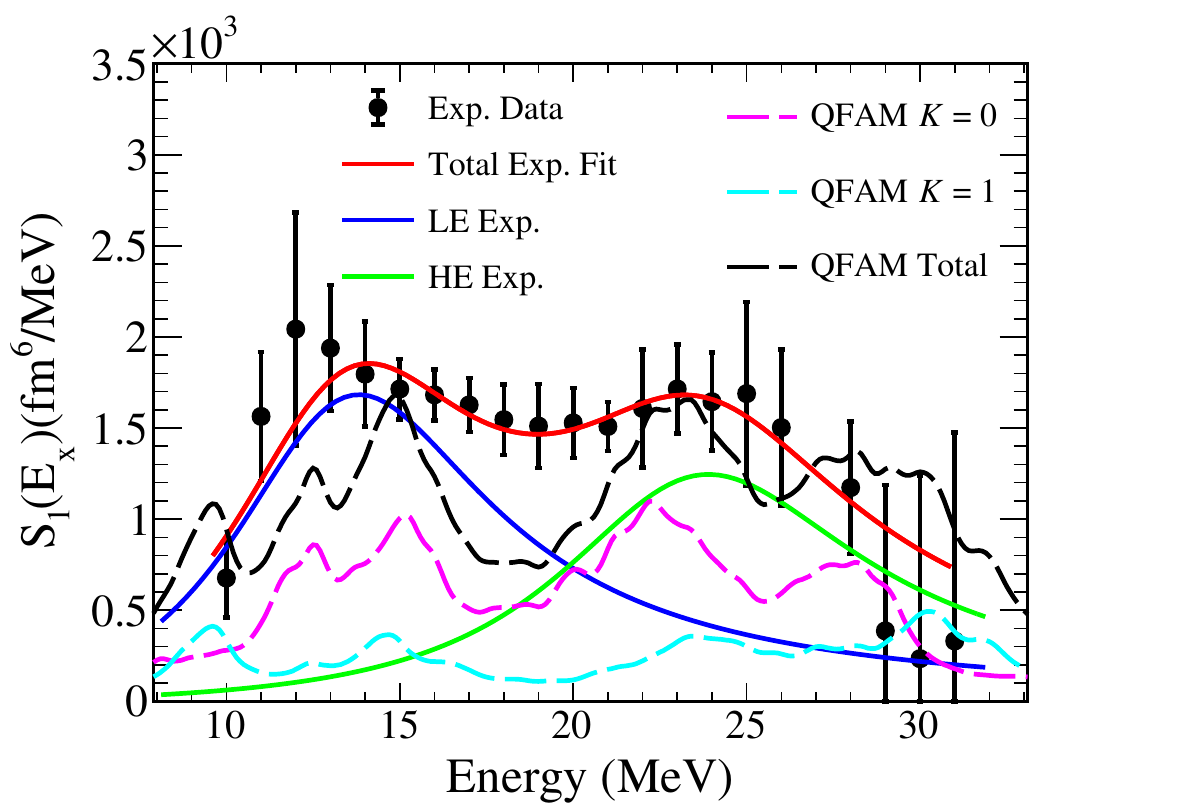}
\caption{Experimental ISGDR strength distribution in $^{172}$Yb extracted from MDA. The error bars on the data points include systematic uncertainties. The solid red line represents the double Lorentzian fit to the experimental data, with the solid blue and green lines representing the LE and HE components, respectively. The black dashed line shows the total ISGDR strength (see text) calculated within the QFAM framework, while the magenta and cyan dashed lines correspond to the $K=0$ and $K=1$ components, respectively. The QFAM strength is multiplied by an additional factor of 4 to compare it with experimental results.}
    \label{ISGDR}
\end{figure}

The transition operator used to extract the isoscalar dipole strength is given by:
\begin{equation}
    O^{1 \mu} = \frac{Z}{2A} \sum_k r_k^{3}Y_{1}^{\mu} (r\hat{}_k).
     \label{Operator_Dipole}
\end{equation}
Here, $Y_{1}^{\mu}$ denotes the spherical harmonics of order $L=1$, with $\mu$ representing the projection of the orbital angular momentum. This corresponds to the second-order (overtone) term in the expansion of the spherical Bessel function, whereas the first-order term describes the spurious center-of-mass translational motion. The ISGDR ($L=1$) strength distribution in $^{172}$Yb was extracted through MDA using the strength formula described in Ref.~\cite{PRC_MA}. The ISGDR strength distribution is shown in Fig.~\ref{ISGDR}. The errors of the data points include the systematic uncertainties discussed in Sec.~\ref{ISGMR_Disc} and the error arising from the MDA fit. The ISGDR strength distribution has two peaks: $1\hbar\omega$ and $3\hbar\omega$. However, the limited acceptance of the spectrometer prevents the measurement of the $1\hbar\omega$ component, which is located below 10 MeV (according to $E_{\rm x} = 30A^{-1/3}$ MeV)~\cite{Poelhekken1992}.

The remaining ISGDR strength distribution shows two distinct peaks, LE and HE, indicating a clear \textit{bimodal} structure as shown in Fig.~\ref{ISGDR}, consistent with earlier observations~\cite{Itoh2003, MUchida2004, Nayak2006}. The LE peak lies above the $1\hbar\omega$ component. The ISGDR strength distribution was fitted with a double Lorentzian function over the 10 to 31 MeV range, shown by the solid red line in Fig.~\ref{ISGDR}, with the LE and HE components represented by solid blue and green lines, respectively. The functional form of the Lorentzian is described in Ref.~\cite{PRC_MA}. The LE component has been linked to toroidal or vortex-like motion~\cite{JKvasil2011, ARepko2013}. The corresponding fitting parameters, including centroid energy $E_{\rm x}$, width $\Gamma$, and the extracted EWSR values for both the LE and HE components, are summarized in Table~\ref{ISGDR_Tab}. The uncertainties in each parameter correspond to the 68$\%$ confidence interval, and the errors in the extracted EWSRs are determined by propagating the uncertainties of the fitted parameters. The EWSR for the LE component (solid blue line) is obtained by integrating $E_{\rm x} S_1(E_{\rm x})$ over 10 to 18 MeV, while for the HE component (solid green line), the integration is performed over 18 to 26 MeV, where $S_1(E_{\rm x})$ was estimated from the Lorentzian fits. An enhancement of the total ISGDR EWSR to about 214$\%$ is observed. Similar enhancements have been reported in other nuclei, such as $\sim$177$\%$ in $^{120,122}$Sn~\cite{TLi2010}, 184$\%$ in $^{100}$Mo~\cite{kevin_thesis2020}, and 198$\%$ in $^{146}$Nd~\cite{PRC_MA}. In addition, deformation effects lead to an increase of the LE strength relative to the HE strength, indicating a shift of strength from HE to LE in deformed nuclei~\cite{Itoh2002, Itoh2003}.

\begin{table*}
\centering
\caption{Experimental fitting parameters of the ISGDR in $^{172}$Yb, including the centroid energy $E_{\rm x}$, width $\Gamma$, and EWSR fractions for the LE and HE components. The uncertainties correspond to the 68$\%$ confidence intervals.}
\renewcommand{\arraystretch}{2.0}
\setlength{\tabcolsep}{21.9pt}
\begin{tabular}{lcccccc}
\hline
\hline
 \multicolumn{3}{c}{LE component} & \multicolumn{3}{c}{HE component} \\
 $E_{\rm x}$ (MeV) & $\Gamma$ (MeV) & EWSR (\%) & $E_{\rm x}$ (MeV) & $\Gamma$ (MeV) & EWSR (\%) \\
\hline
 $13.8 \pm 0.3$ & $9.1 \pm 0.7$ & $99.9^{+7.2}_{-7.8}$ & $23.9 \pm 0.5$ & $10.8 \pm 1.3$ & $114.2^{+8.6}_{-9.8}$ \\
\hline
\hline
\label{ISGDR_Tab}
\end{tabular}
\end{table*}

A discrepancy of $Z/2A$ exists between the dipole transition operator employed in the present experimental analysis (see Eq.~\ref{Operator_Dipole}) and that used in the theoretical calculations within the QFAM framework for the ISGDR (see Eq.~\ref{QFAM_Op}). By scaling the QFAM strength distribution with a factor of $Z^2/4A^2$, the resulting expression corresponds to the operator form as Eq.~\eqref{Operator_Dipole}, consistent with the strength formula reported in Ref.~\cite{PRC_MA}. However, an additional factor of 4 has been multiplied to the QFAM strength to compare with the experimental results. The theoretical ISGDR strength distribution calculated within the QFAM framework is shown by the black dashed line in Fig.~\ref{ISGDR}. This total strength is obtained as the sum of the $K=0$ and $2\times(K=1)$ components, where the factor of 2 for the $K=1$ component accounts for its time-reversal partner ($K=-1$). The components $K=0$ and $K=1$ are represented by magenta and cyan dashed lines, respectively, in Fig.~\ref{ISGDR}. The experimental strength shows overall agreement with the QFAM results, although QFAM predicts additional splitting that lies beyond the experimental resolution.

The widths of both the LE and HE components are known to increase with nuclear deformation~\cite{Itoh2003, Yoshida2013}, as the ISGDR couples with the $K=0$ and $K=1$ components of the HEOR~\cite{Yoshida2013}. As the deformation of $^{172}$Yb is comparable to that of $^{154}$Sm, the LE width in $^{172}$Yb is consistent with that of $^{154}$Sm~\cite{Itoh2003}, whereas the HE width is larger in $^{154}$Sm. Notably, the width of the HE component in the experimental distribution for $^{172}$Yb is comparable to the QFAM prediction at around 25 MeV.


\subsection{ISGQR}
\label{ISGQR_Disc}
The transition operator used to extract the multipole ($L \geq 2$) strengths is given as
\begin{equation}
 O^{L \mu}=\frac{Z}{A} \sum_k r_k^{L}Y_{L}^{\mu} (\hat{r}_k).
 \label{Operator_quadrupole_and_higher}
\end{equation}

\begin{figure}[hbt!]
    \centering
    \includegraphics[width=1.1\linewidth]{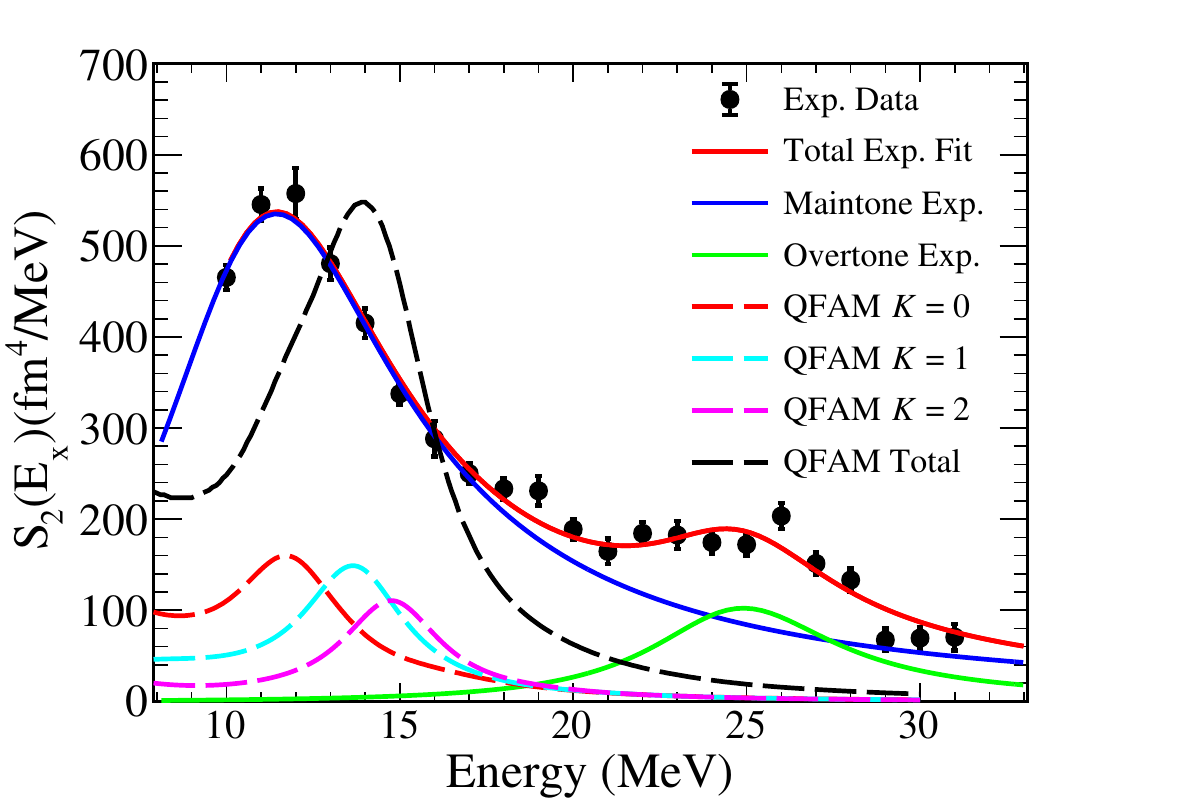}
    \caption{Experimentally extracted ISGQR strength distribution in $^{172}$Yb obtained through MDA. The error bars on the data points include the systematic uncertainties. The experimental main-tone and overtone components are shown by the solid blue and solid green lines, respectively. The theoretical ISGQR total strength distribution (see text) calculated within the QFAM framework is represented by the black dashed line. The QFAM $K$ components are shown as follows: $K=0$ (red dashed line), $K=1$ (cyan dashed line), and $K=2$ (magenta dashed line).}
    \label{ISGQR}
\end{figure}

$Y_{L}^{\mu}$ is the spherical harmonic of order $L$, with $\mu$ as its projection. This term represents the main-tone (first-order) component of the spherical Bessel function. For the ISGQR, it has an energy of $2\hbar\omega$, unlike the monopole and dipole modes. The ISGQR ($L=2$) strength distribution for $^{172}$Yb was extracted using MDA, and the corresponding strength was evaluated with the strength formula given in Ref.~\cite{PRC_MA}. The experimental ISGQR strength distribution in $^{172}$Yb is presented in Fig.~\ref{ISGQR}. The uncertainties in the data points include both systematic errors, as explained in Sec.~\ref{ISGMR_Disc}, and fitting errors arising from the MDA. 

The strength distribution is fitted with a double Lorentzian function in the energy range of 10 to 30 MeV, as shown by the solid red line in Fig.~\ref{ISGQR}. The corresponding fitting parameters are provided in Table~\ref{ISGQR_Tab}. The uncertainties in all the fitting parameters fall within the 68$\%$ confidence interval, and the uncertainty in the EWSR is calculated based on the errors of the fitting parameters. The main-tone mode of the ISGQR strength, shown by the solid blue line in Fig.~\ref{ISGQR}, corresponds to the operator defined in Eq.~\eqref{Operator_quadrupole_and_higher} and is associated with the $2\hbar\omega$ excitation energy. In deformed nuclei, this main-tone mode splits into three $K$ components $K=0, 1$, and $2$. These components are closely spaced, leading to an increase in the overall ISGQR width in deformed nuclei~\cite{Yoshida2013}. Although $^{172}$Yb and $^{154}$Sm have similar deformations, the width of the main-tone component is larger in $^{172}$Yb than in $^{154}$Sm~\cite{Itoh2003}. The theoretical ISGQR strength, calculated within the QFAM framework using the operator $\sum_k r_k^{2}Y_{2}$, corresponds, therefore, only to the main-tone mode. A discrepancy arises due to the $Z/A$ factor between the dipole transition operator employed in the present experimental analysis (see Eq.~\ref{Operator_quadrupole_and_higher}) and that used in the theoretical calculation within the QFAM framework for the ISGQR (see Eq.~\ref{QFAM_Op}). To resolve this, the QFAM ISGQR strength is scaled by a factor of $Z^2/A^2$, ensuring that the resulting strength corresponds to the operator defined in Eq.~\eqref{Operator_quadrupole_and_higher} and remains consistent with the strength formulas reported in Ref.~\cite{PRC_MA}. The scaled QFAM strength is plotted as the black dashed line in Fig.~\ref{ISGQR}. The corresponding $K$ components of the QFAM ISGQR strength distribution are displayed as red ($K=0$), cyan ($K=1$), and magenta ($K=2$) dashed lines. The total ISGQR strength in QFAM is obtained as $(K=0) + 2\times(K=1) + 2\times(K=2)$, where the factors of 2 account for the time-reversal partners $K=-1$ and $K=-2$ of the components $K=1$ and $K=2$, respectively. The QFAM strength has not been evaluated for the $\sum_k r_k^{4}Y_{2}$ operator; consequently, the overtone mode is absent in the QFAM results. The experimental ISGQR main-tone strength shows overall agreement with the QFAM results. Furthermore, the experimental ISGQR main-tone centroid energy is consistent with the LE centroid energy of the ISGMR within the 1 MeV energy binning.

\begin{table*}[hbt!]
\centering
\caption{Experimental fitting parameters of the ISGQR strength in $^{172}$Yb. The centroid energy $E_{\rm x}$ and width $\Gamma$ for the main-tone and overtone components, together with the corresponding extracted EWSR values, are listed. The quoted uncertainties correspond to 68$\%$ confidence interval.}
\renewcommand{\arraystretch}{2.0}
\setlength{\tabcolsep}{21.9pt}
\begin{tabular}{lcccccc}
\hline
\hline
 \multicolumn{3}{c}{Main-tone} & \multicolumn{3}{c}{Overtone} \\
 $E_{\rm x}$ (MeV) & $\Gamma$ (MeV) & EWSR (\%) & $E_{\rm x}$ (MeV) & $\Gamma$ (MeV) & EWSR (\%) \\
\hline
 $11.4 \pm 0.1$ & $8.5 \pm 0.1$ & $78.1 \pm 2.4$ & $24.9 \pm 0.3$ & $6.5 \pm 0.6$ & $26.3^{+2.7}_{-2.8}$ \\
\hline
\hline
\label{ISGQR_Tab}
\end{tabular}
\end{table*}

An enhancement of the experimental ISGQR strength is observed around 25 MeV, which is attributed to the overtone mode corresponding to the $4\hbar\omega$ frequency. The experimental strength distributions for both the main-tone and overtone modes were calculated using the operator defined in Eq.~\eqref{Operator_quadrupole_and_higher}. In contrast, the overtone strength distribution reported in Refs.~\cite{Gorelik2004, Gorelik2021, Gorelik2023}, obtained with the operator $r^{4}-\eta_{L} r^{2}$, exhibits a broader peak shifted to higher excitation energies relative to the present results. The parameter $\eta_{L}$ is dependent on the multipolarity of the mode. Evidence for the ISGQR overtone has previously been reported in $^{208}$Pb~\cite{Hunyadi2003, Hunyadi2007}, while further indications were observed in $^{90,92}$Zr and $^{92}$Mo, where an enhancement of the ISGQR strength above 20 MeV was identified~\cite{YKGupta2018}. The overtone mode was identified for the first time through MDA in Nd isotopes~\cite{Abdullah2024}. 
The EWSR for the main-tone and overtone modes is obtained by integrating $E_{\rm x} S_2(E_{\rm x})$ in the excitation-energy ranges 10 to 20 MeV and 20 to 30 MeV, respectively as shown in Table~\ref{ISGQR_Tab}, where $S_2(E_{\rm x})$ was estimated from the Lorentzian fits.


\subsection{HEOR}
\label{HEOR_Disc}

\begin{figure}[hbt!]
    \centering
    \includegraphics[width=1.0\linewidth]{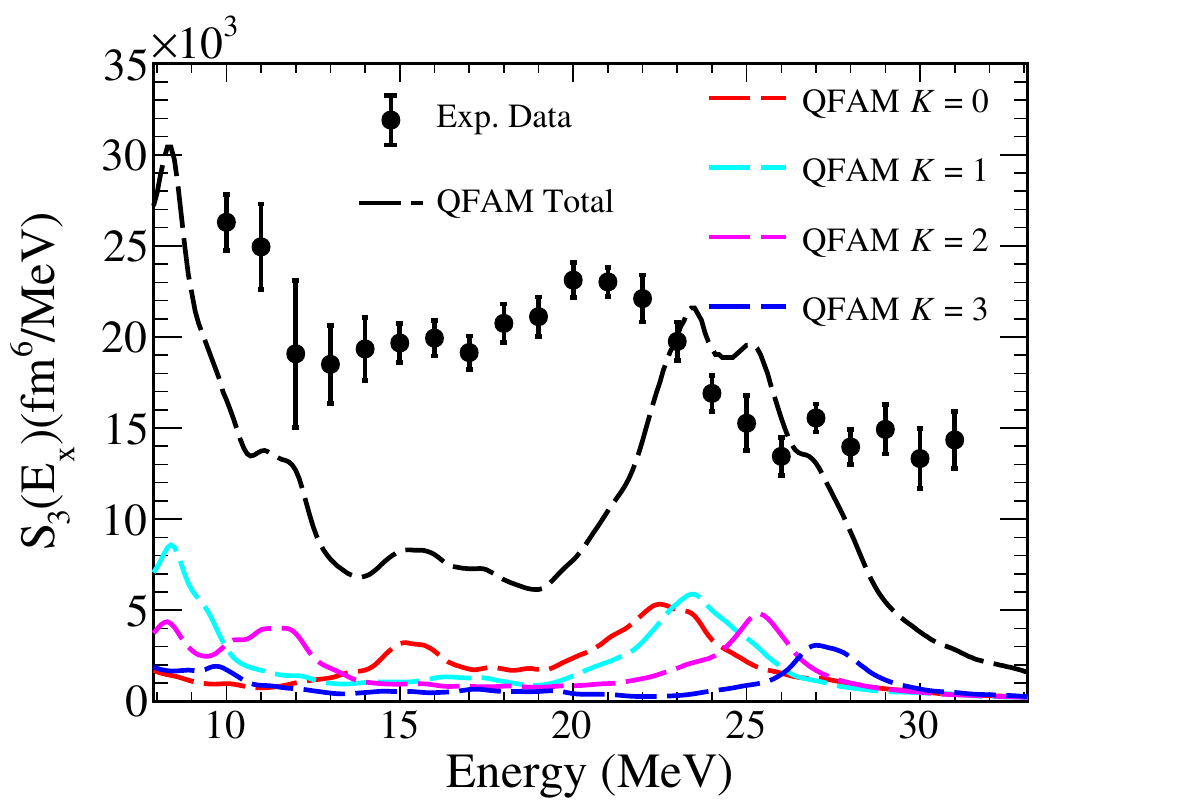}
    \caption{The HEOR strength distribution in $^{172}$Yb is shown with the data point uncertainties including systematic errors. The full energy range of the HEOR, from 10 to 31 MeV, is divided into two regions: LE (10 to 17 MeV) and HE (17 to 31 MeV). The total QFAM strength (see text) is shown as a black dashed line, while the corresponding $K$ components $K = 0, 1, 2, 3$ are displayed as red, cyan, magenta, and blue dashed lines, respectively.}
    \label{ISHEOR}
\end{figure}

The HEOR ($L=3$) strength distribution in $^{172}$Yb was extracted through MDA. The transition operator for the octupole mode is defined in Eq.~\ref{Operator_quadrupole_and_higher}, and the HEOR strength is calculated using the formula defined in Ref.~\cite{PRC_MA}. The resulting HEOR strength distribution is presented in Fig.~\ref{ISHEOR}, where uncertainties in the data points include both systematic errors as defined in Sec.~\ref{ISGMR_Disc} and the errors arising from the MDA fitting procedure. The low-energy octupole resonance, expected at $E_{\rm x} \approx 30A^{-1/3}$ MeV~\cite{Itoh2003, Morsch1982}, lies below the acceptance of the GR spectrometer. The observed HEOR strength spans the region from 10 to 31 MeV and is divided into two components: an LE component between 10 to 17 MeV and an HE component between 17 to 31 MeV, as discussed for $^{154}$Sm~\cite{Itoh2003} and the Nd isotopes~\cite{PRC_MA}. Centroid energies for the LE and HE components were evaluated as the average excitation energies within the respective energy intervals. Widths are not quoted here, as a Lorentzian fit of the HEOR strength distribution was not performed. The $\%$EWSR fractions were obtained by summing the EWSR fraction for each energy bin over 10 to 17 MeV for the LE component and 17 to 31 MeV for the HE component. The extracted centroid energies and EWSR values are summarized in Table~\ref{ISHEOR_Tab}. The $\%$EWSR values for the LE and HE components are larger than those reported for $^{154}$Sm~\cite{Itoh2003}. This difference arises from the MDA, where multipolarities up to $L_{\text{max}}=9$ were included, compared to $L_{\text{max}}=12$ in the $^{154}$Sm analysis. The use of $L_{\text{max}}=7$ in the MDA of Nd isotopes resulted in even higher $\%$EWSR values for both the LE and HE components~\cite{PRC_MA}. In determining the OMPs as discussed in Sec.~\ref{OMP_Disc}, the partial wave number in CHUCK3 was fixed at 300, as lower values gave poor fits to the elastic-scattering data. This allowed angular distributions up to $L=9$, but CHUCK3 limitations require reducing the partial wave number for $L>9$. To maintain consistency, the analysis was restricted to below $L=10$. 

The HEOR strength has been theoretically calculated within the QFAM framework using the operator defined in Eq.~\ref{QFAM_Op} and compared with the experimental results. The QFAM HEOR strength was scaled by a factor of $Z^2/A^2$, following the procedure described in Sec.~\ref{ISGQR_Disc}. The scaled QFAM HEOR strength is shown as a black dashed line in Fig.~\ref{ISHEOR}. The corresponding $K$ components $K = 0, 1, 2, 3$ of the QFAM strength are also plotted in Fig.~\ref{ISHEOR} as red, cyan, magenta, and blue dashed lines, respectively. The total QFAM strength was obtained as $(K=0)+2\times(K=1)+2\times(K=2)+2\times(K=3)$, where the factor of 2 for $K = 1, 2, 3$ represents the time-reversal partners with $K = -1, -2, -3$, respectively. The overall trend of the QFAM strength distribution agrees reasonably well with the experimental data. While the calculated strength is somewhat lower in the low-energy region, good agreement is observed in the high-energy region. The centroid energies of the LE and HE components of the HEOR obtained experimentally are consistent with the QFAM result.

\begin{table}[hbt!]
\centering
\caption{The centroid energies ($E_{\rm x}$) and EWSR values for the LE and HE components of the HEOR are listed. The method used for their extraction is described in Sec.~\ref{HEOR_Disc}.}
\renewcommand{\arraystretch}{2.0}
\setlength{\tabcolsep}{8.9pt}
\begin{tabular}{ccccc}
\hline
\hline
\multicolumn{2}{c}{LE} & \multicolumn{2}{c}{HE} \\
$E_{\rm x}$ (MeV) & EWSR (\%) & $E_{\rm x}$ (MeV) & EWSR (\%)\\
\hline
$14.2 \pm 0.7$ & $33.9 \pm 2.9$ & $23.7 \pm 1.1$ & $90.0\pm 6.2$ \\
\hline
\hline
\label{ISHEOR_Tab}
\end{tabular}
\end{table}


\section{Summary}
Inelastic $\alpha$-particle scattering on $^{172}$Yb was measured with the Grand Raiden spectrometer at $0^{\circ} \leq \theta_{\text{Lab}} \leq 10^{\circ}$ to study the isoscalar giant resonances. After particle identification, instrumental background subtraction, ion-optical corrections, removal of hydrogen contamination, and subtraction of the IVGDR contribution, the data were subjected to MDA to extract the strength distributions. The OMPs used in the DWBA calculations for different multipoles were obtained from fitting elastic-scattering data with the Woods–Saxon potential. Strength distributions were extracted for multipoles up to $L=3$ and are discussed in their respective sections. Theoretical strength calculations were performed within the QFAM framework up to $L=3$.

A clear splitting of the ISGMR strength distribution has been observed, attributed to the coupling of the ISGMR with the $K=0$ component of the ISGQR. The experimental data show good agreement with QFAM predictions for both the LE and HE components, within the 1 MeV energy binning. Furthermore, the HE component is also consistent with QRPA results~\cite{JKvasil2016}. The LE component of the QRPA is below 10 MeV, which is beyond the acceptance of the spectrometer.

The ISGDR strength distribution exhibits the characteristic \textit{bimodal} structure. The experimental strength shows overall good agreement with the QFAM predictions. The enhanced experimental LE component strength indicates a shift of strength from HE to LE. This shift is interpreted as a consequence of coupling between the $K=0$ and $K=1$ components of the ISGDR and HEOR in $^{172}$Yb.

The ISGQR strength distribution displays a broad main-tone component, attributed to deformation effects in $^{172}$Yb. The QFAM calculations for the ISGQR were performed using the operator $\sum_k r_k^{2}Y_{2}$; therefore, only the main-tone component is predicted and compared with the experimental data. An enhancement in the experimental ISGQR strength around 25 MeV is observed, which is consistent with the expected overtone mode, representing a third type of compression mode in addition to the ISGMR and ISGDR. Evidence of the ISGQR overtone was reported in $^{208}$Pb~\cite{Hunyadi2003, Hunyadi2007}, with further indications in $^{90,92}$Zr and $^{92}$Mo from enhanced strength above 20 MeV~\cite{YKGupta2018}. The overtone mode was identified for the first time in Nd isotopes through MDA~\cite{Abdullah2024}. The QFAM calculation has not been carried out with the operator $\sum_k r_k^{4}Y_{2}$; hence, the overtone peak is absent in the QFAM results.

The experimental HEOR strength extends over the 10 to 31 MeV energy range and is divided into LE and HE components, with the corresponding EWSR fractions obtained by summing over their respective energy intervals. The experimental HEOR strength is compared with the theoretical results calculated within the QFAM framework. QRPA calculations are not available for modes other than the ISGMR~\cite{JKvasil2016}.


\section{Acknowledgments}
The authors gratefully acknowledge the staff of the RCNP Ring Cyclotron Facility for providing a high-quality, halo-free $\alpha$ beam and for their excellent local support during the experiment. The self-supporting $^{172}$Yb target used in this work was kindly provided by RCNP. S.B. acknowledges financial support from the Science and Engineering Research Board (SERB), now reorganized as the Anusandhan National Research Foundation (ANRF), India (Grant No. SRG/2021/000827), and from the Faculty Research Scheme at IIT (ISM) Dhanbad (Grant No. FRS(154)/2021-2022/Physics). Additional support from GSI Helmholtzzentrum, Germany, in the form of travel assistance for participation in the experiment is gratefully acknowledged. K.K. thanks IIT (ISM) Dhanbad for the institute fellowship. This work was partially supported by the U.S. National Science Foundation (Grants No. PHY-1713857 and PHY-2310059).

\bibliography{apssamp}

\end{document}